     \documentstyle[12pt,epsfig,cite,amssymb]{article}
     \newlength{\dinwidth}
     \newlength{\dinmargin}
\def\Journal#1#2#3#4{{#1} {\bf #2}, #3 (#4)}

\def\NPB{{\em Nucl. Phys.} B}
\def\PLB{{\em Phys. Lett.}  B}

\def\ZPC{{\em Z. Phys.} C}
\def\lsim{\mathrel{\rlap{\lower4pt\hbox{\hskip1pt$\sim$}}
    \raise1pt\hbox{$<$}}}                
\def\gsim{\mathrel{\rlap{\lower4pt\hbox{\hskip1pt$\sim$}}
    \raise1pt\hbox{$>$}}}                
\begin{document}
\vspace*{10mm}
\begin{center}  \begin{Large} \begin{bf}
Interpreting virtual photon interactions in terms of parton
distribution functions
\footnote{Contribution to the Workshop {\em Monte Carlo generators for
HERA Physics}, DESY, 1998--1999}
\\
  \end{bf}  \end{Large}
  \vspace*{5mm}
  \begin{large}
J. Ch\'{y}la$^a$, M. Ta\v{s}evsk\'{y}$^{a}$\\
  \end{large}
\end{center}
$^a$ Institute of Physics, Prague, Na Slovance 8, Prague 8, Czech
Republic\\
\begin{quotation}
\noindent
{\bf Abstract:} Interactions of virtual photons are analyzed in
terms of photon structure. It is argued that the concept of parton
distribution functions is phenomenologically very useful even for
highly virtual photons involved in hard collisions. The role of
longitudinal photons for proper interpretation of jet
cross--sections in the region of moderate virtualities accessible at
HERA is investigated.
\end{quotation}
\section{Introduction}
In quantum field theory it is difficult to distinguish effects of
the ``structure'' from those of ``interactions''. Within the
Standard Model it makes good sense to distinguish {\em fundamental
particles}, which correspond to fields in its lagrangian ${\cal
L}_{\mathrm{SM}}$ (leptons, quarks, gauge and Higgs bosons) from
{\em composite particles}, like atoms or hadrons, which appear in
the mass spectrum but have no corresponding field in ${\cal
L}_{\mathrm{SM}}$. For the latter the use of parton distribution
functions (PDF) to describe their ``structure'' appears natural,
but the concept of PDF turns out to be phenomenologically useful
also for some fundamental particles, in particular the photon. PDF
are indispensable for the real photon due to strong interactions
between the pair of quarks to which it couples electromagnetically.
For massless quarks this coupling leads to parallel singularities,
which must be absorbed in PDF of the photon. For nonzero photon
virtualities there is no true singularity associated with the
coupling $\gamma\rightarrow q\overline{q}$ and therefore no real
need for PDF. The main aim of this paper is to advocate the use of
PDF even for the virtual photon.

\section{PDF of the real photon}
\begin{figure}\centering \unitlength 1mm
\begin{picture}(120,40)
\put(0,20){\epsfig{file=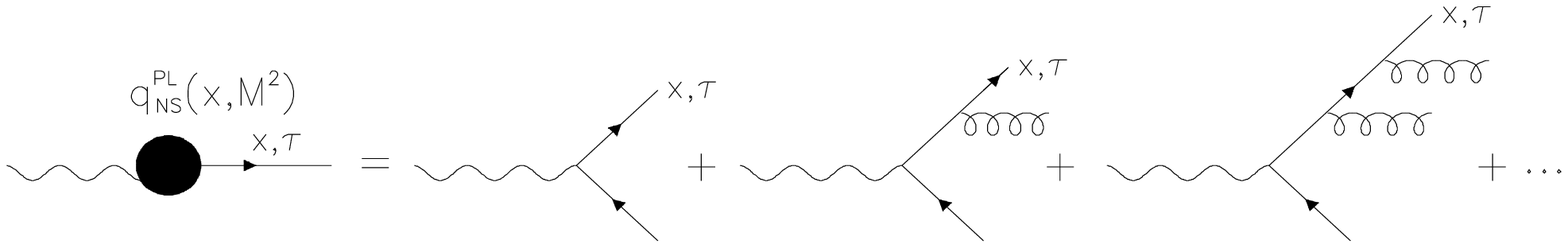,width=12cm}}
\put(0,0){\epsfig{file=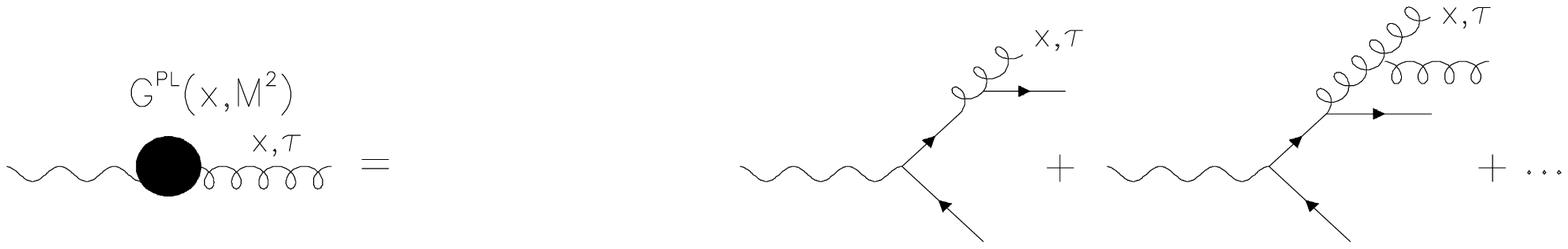,width=12cm}}
\end{picture}
\caption{\em Diagrams defining the pointlike parts of
quark and gluon distribution functions in LL approximation. The
resummation involves integration over parton virtualities $M_0^2\le
\tau\le M^2$.}
\label{figpl}
\end{figure}

The factorization scale dependence of PDF of the real photon is
determined by the system of coupled inhomogeneous evolution
equations
\begin{eqnarray}
\frac{{\mathrm d}\Sigma(M^2)}{{\mathrm d}\ln M^2}& =&
k_q+P_{qq}\otimes \Sigma+ P_{qG}\otimes G,
\label{Sigmaevolution} \\
\frac{{\mathrm d}G(M^2)}{{\mathrm d}\ln M^2} & =&
k_G+ P_{Gq}\otimes \Sigma+ P_{GG}\otimes G,
\label{Gevolution} \\
\frac{{\mathrm d}q_{\mathrm NS}(M^2)}{{\mathrm d}\ln M^2}& =&
\sigma_{\mathrm NS} k_q+P_{\mathrm NS}\otimes q_{\mathrm NS},
\label{NSevolution}
\end{eqnarray}
for singlet, nonsinglet and gluon distribution functions. The
spliting functions $k_q,k_G$ and $P_{ij}$ admit expansions in
powers of $\alpha_s$,
\begin{eqnarray}
k_q(x,M) & = & \frac{\alpha}{2\pi}\left[k^{(0)}_q(x)+
\frac{\alpha_s(M)}{2\pi}k_q^{(1)}(x)+
\left(\frac{\alpha_s(M)}{\pi}\right)^2k^{(2)}_q(x)+\cdots\right],
\label{splitquark} \\
k_G(x,M) & = & \frac{\alpha}{2\pi}\left[~~~~~~~~~~~~
\frac{\alpha_s(M)}{2\pi}k_G^{(1)}(x)+
\left(\frac{\alpha_s(M)}{\pi}\right)^2k^{(2)}_G(x)+\cdots\right],
\label{splitgluon} \\
P_{ij}(x,M) & = &
~~~~~~~~~~~~~~~~~~\frac{\alpha_s(M)}{2\pi}P^{(0)}_{ij}(x) +
\left(\frac{\alpha_s(M)}{2\pi}\right)^2 P_{ij}^{(1)}(x)+\cdots,
\label{splitpij}
\end{eqnarray}
which start at the order $\alpha\alpha_s$, except for $k_q$ for
wich $k_q^{(0)}=3e_q^2(x^2+(1-x)^2)=O(1)$. The general solution of
these equations can be written as the sum of a particular solution
of the full inhomogeneous equations and the general solution of the
corresponding homogeneous ones, called {\em hadronic} (or VDM)
part. A subset of solutions of the inhomogeneous evolution
equations resulting from the resummation of diagrams in Fig.
\ref{figpl} defines the so called {\em pointlike} (PL) parts. This
resummation softens the $x-$dependence of $q^{\mathrm {PL}}(x,M^2)$
with respect to the term $(\alpha/2\pi)k_{\mathrm {NS}}^{(0)}(x)$,
corresponding to the simple $\gamma\rightarrow q\overline{q}$
splitting. A general solution of eqs.
(\ref{Sigmaevolution}-\ref{Gevolution}) can thus be written as
($D=q,\overline{q},G$)
\begin{equation}
D(x,M^2)= D^{\mathrm {PL}}(x,M^2)+D^{\mathrm {VDM}}(x,M^2),
~~D=q,\overline{q},G.
\label{separation}
\end{equation}
The important point to keep in mind is the fact that there is an
infinite number of pointlike solutions $q^{\mathrm {PL}}(x,M^2)$,
$G^{\mathrm {PL}}(x,M^2)$, differing merely by the initial scale
$M_0$ at which they vanish. Consequently, the separation of quark
and gluon distribution functions into their pointlike and VDM parts
is ambiguous and these concepts have thus separately no physical
meaning
\footnote{For brevity the terms ``pointlike quarks''
and ``pointlike gluons'' will hence be employed to denote pointlike
parts of quark and gluon distribution functions of the photon.}.

\subsection{Properties of Schuler--Sj\"{o}strand parametrizations}
\begin{figure}\centering
\epsfig{file=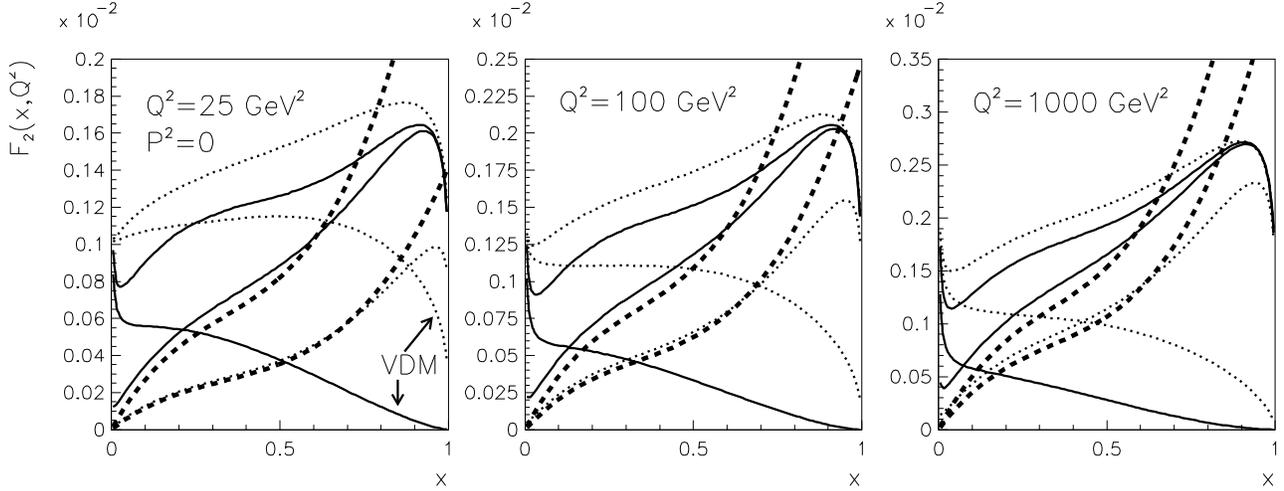,width=\textwidth}
\caption{\em $F_2^{\gamma}(x,Q^2)$ as a function of $x$ for $Q^2=25,
100,1000$ GeV$^2$ as given by SaS1D (solid curves) and SaSD2
(dotted curves) parametrizations. The full results are given by the
upper, the VDM and pointlike contributions parts by two lower
curves. The dashed curves describe the contributions of the
splitting term (\ref{splitterm}) with $M_0=0.6$ GeV for SaS1D and
$M_0=2$ GeV for SaS2D.}
\label{f2dreal}
\end{figure}

\begin{figure}\centering
\epsfig{file=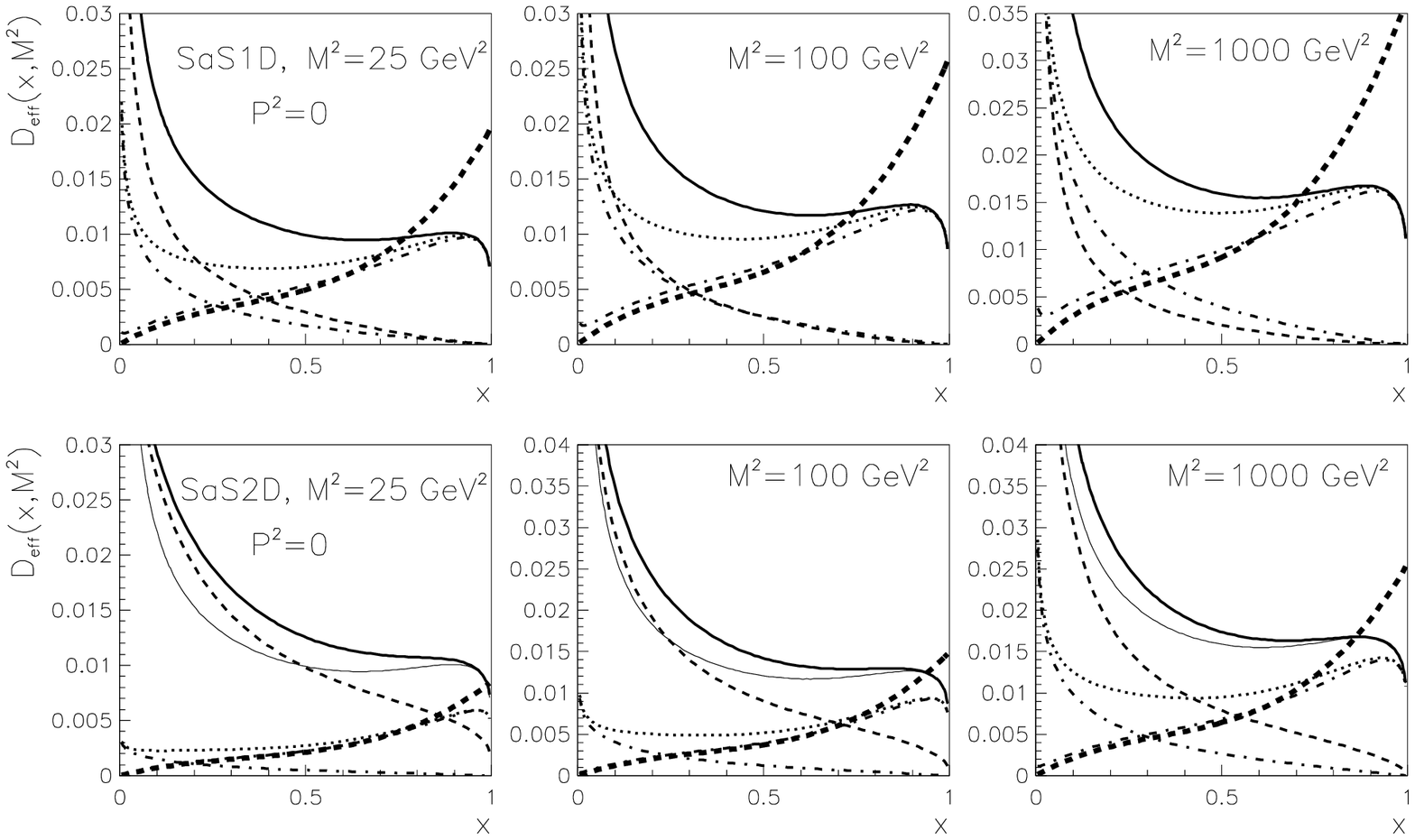,width=\textwidth}
\caption{\em $D_{\mathrm eff}(x,M^2)$ as a function of $x$ for $M^2=25,
100,1000$ GeV$^2$ for the real photon as given by SaS1D and SaSD2
parametrizations. Solid curves show the full results, dashed ones
the VDM and dotted ones the pointlike parts. The pointlike part is
further separated into the contributions of pointlike quarks and
pointlike gluons, denoted by two dash--dotted curves (those peaking
at low $x$ correspond to gluons). Thick dashed curves correspond to
the splitting term (\ref{splitterm}). The thin solid curves in
SaS2D plots show the full results of the SaS1D parametrization.}
\label{eff3dreal}
\end{figure}

Practical aspects of the ambiguity in separating PDF into their VDM
and pointlike parts can be illustrated on the properties of SaS1D
and SaS2D parametrizations \cite{sas1,sas2},
which differ by the choice of initial $M_0$: $M_0=0.6$
GeV for SaS1D, $M_0=2$ GeV for SaS2D. What makes the SaS approach
particularly useful for our discussion is the fact that it provides
separate parametrizations of the VDM and pointlike parts of both
quark and gluon distributions.
\begin{figure}[t]\centering
\epsfig{file=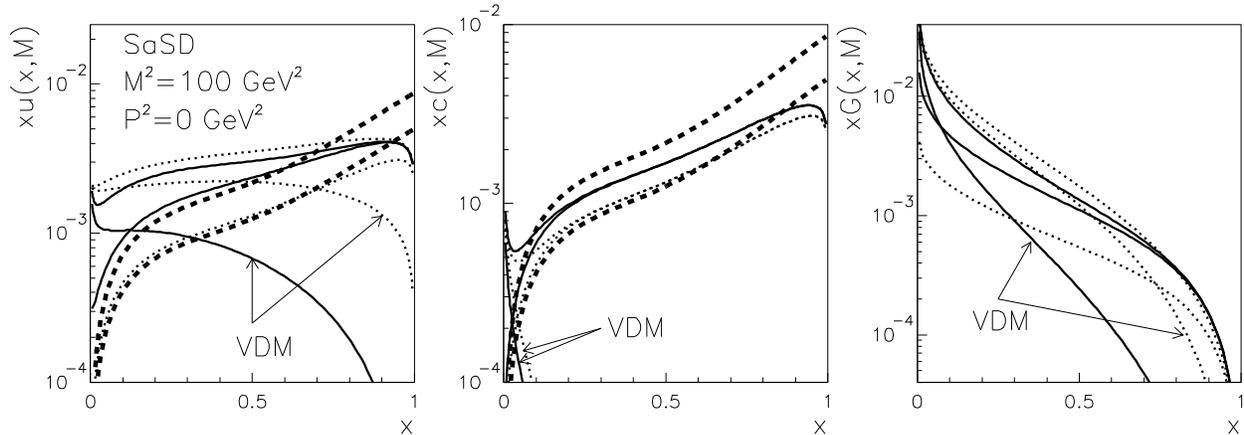,width=\textwidth}
\caption{\em The $u$ and $c$ quark and gluon distribution
functions of the real photon for SaS1D (upper solid curves) and
SaS2D (upper dotted curves) parametrizations at $M^2=100$ GeV$^2$.
The VDM and pointlike (solid and dotted curves peaking at large
$x$) parts of both parametrizations are plotted separately. For
quarks the splitting terms corresponding to SaS1D (upper dashed
curves) and SaS2D (lower dashed curves) parametrizations are
overlayed.}
\label{sasd1real}
\end{figure}
In Fig. \ref{sasd1real} we compare distribution functions
$xu(x,M^2)$, $xc(x,M^2)$ and $xG(x,M^2)$ as given by SaS1D and
SaS2D parametrizations at $M^2=100$ GeV$^2$. To see the effects of
the resummation of multiple parton emission we also plot the
corresponding splitting terms
\begin{equation}
q^{\mathrm split}(x,M_0^2,M^2)\equiv
\frac{\alpha}{2\pi}3e_q^2\left(x^2+(1-x)^2\right)\ln\frac{M^2}{M_0^2}.
\label{splitterm}
\end{equation}
In Fig. \ref{scaledependence} we illustrate in two ways the scale
dependence of VDM and pointlike parts of the same three
distribution functions. In the upper six plots we compare them as a
function of $x$ at $M^2=25,100,1000$ GeV$^2$, while in the lower
three plots the same distributions are rescaled by the factor
$\ln(M^2/M_0^2)$. Finally, in Figs. \ref{f2dreal} and
\ref{eff3dreal} we compare LO SaS predictions for two physical
quantities: $F_2^{\gamma}$ and effective parton distribution
function $D_{\mathrm {eff}}$, relevant for approximate calculations
of jet production
\begin{eqnarray}
F_2^{\gamma}(x,Q^2)& = & \sum_{i=1}^{n_f}2x e_i^2 q_i(x,Q^2),
\label{F2PM}\\
D_{\mathrm eff}(x,M^2) & = &
\sum_{i=1}^{n_f}\left(q_i(x,M^2)+\overline{q}_i(x,M^2)\right)+
\frac{9}{4}G(x,M^2).
\label{deff}
\end{eqnarray}
Figures \ref{sasd1real}--\ref{eff3dreal} illustrate several
important properties of PDF of the real photon:
\begin{figure} \unitlength 1mm
\begin{picture}(160,160)
\put(0,50){\epsfig{file=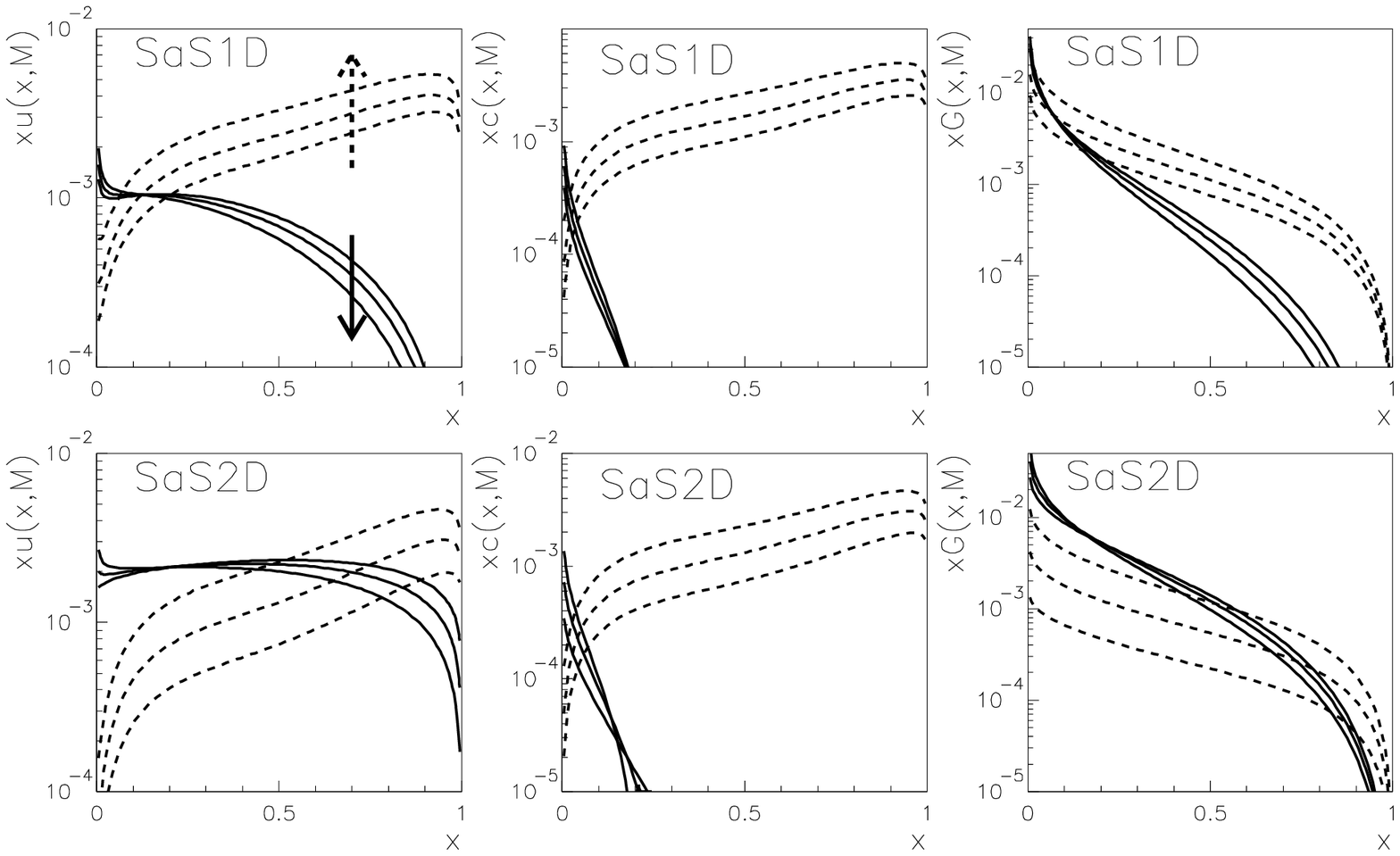,width=\textwidth}}
\put(0,0){\epsfig{file=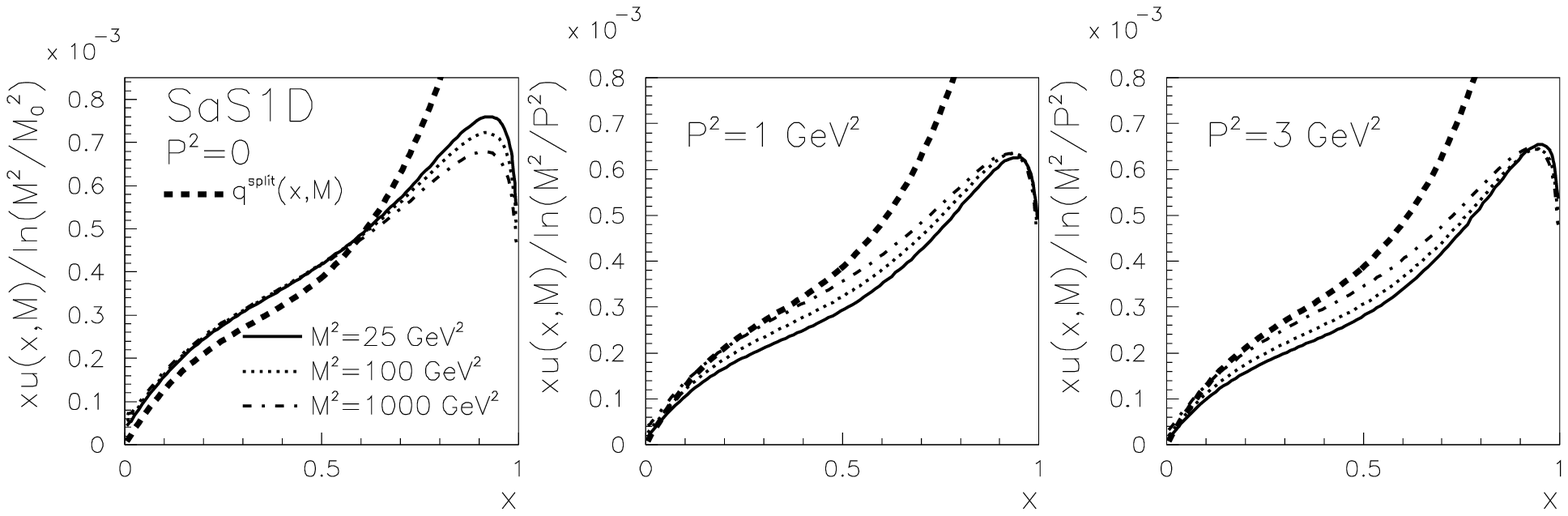,width=\textwidth}}
\end{picture}
\caption{\em Factorization scale dependence of parton distributions
functions $u(x,M),c(x,M)$ and $G(x,M)$ of the real photon. Dashed
curves correspond, in the order indicated by the arrows, to
pointlike and solid to VDM parts of these distributions at
$M^2=25,100$ and $1000$ GeV$^2$. The meaning of arrows is the same
for all parton distribution functions. In the lower part SaS1D
quark distribution functions $xu(x,M^2,P^2)$ scaled by
$\ln(M^2/M_0^2)$ for the
real photon (left) and by $\ln(M^2/P^2)$ for the virtual one, are
plotted and compared to the predictions of the splitting term
(\ref{splitterm}).}
\label{scaledependence}
\end{figure}

\begin{itemize}
\item There is a huge difference between the relative importance
of VDM components in SaS1D and SaS2D parametrizations: for SaS2D
the VDM components of $xu(x,M)$ and $xG(x,M)$ are dominant up to
$x\doteq 0.75$, whereas for SaS1D the pointlike one dominates
already above $x\doteq 0.1$!
\item Factorization scale dependence of VDM and
pointlike parts differ substantially. VDM components exhibit the
pattern of scaling violations typical for hadrons, whereas the
pointlike parts of both quark and gluon distribution functions rise
with $M$ for all $x$. For pointlike gluons this holds despite the
fact that $G^{\mathrm {PL}}(x,M^2)$ satisfies at the LO standard
homogeneous evolution equation and is due to the fact that its
evolution of is driven by the rise of $\Sigma^{\mathrm
{PL}}(x,M^2)$.
\item As the factorization scale $M$ increases the VDM parts of both
quark and gluon distribution functions decrease relative to the
pointlike ones, except in the region of very small $x$.
\item Despite huge differences between SaS1D and SaS2D
parametrizations in the decomposition of quark and gluon
distributions into their VDM and pointlike parts, their predictions
for physical quantities $F_2^{\gamma}$ and $D_{\mathrm {eff}}$ are
quite close.
\item The most prominent effect of multiple parton emission on
physical quantities appears to be the contribution of pointlike
gluons to jet cross--sections in the region $x_{\gamma}\lesssim
0.5$.
\end{itemize}

\section{PDF of the virtual photon: do we really need them?}
For sufficiently virtual photon the initial state singularity
resulting from the splitting $\gamma^*\rightarrow q\overline{q}$ is
shielded off by the nonzero initial photon virtuality $P^2$ and
therefore in principle the concept of PDF does not have to be
introduced. Nevertheless, even in such circumstances PDF turn out
to be phenomenologically very useful because their pointlike parts
include the resummation of parts of higher order QCD corrections,
and the VDM parts, though decreasing rapidly with increasing $P^2$,
are still dominant at very small $x_{\gamma}$. Both of these
aspects define the ``nontrivial'' structure of the virtual photon
in the sense that they are not included in the splitting term
(\ref{splitterm}) and thus are not part of existing NLO
unsubtracted direct photon calculations.

\subsection{Virtuality dependent PDF}
In QCD the nonperturbative effects connected with the confinement
are expected to determine the long--range structure of the photon
and hence also the transition between the virtual and real photon.
As for the real photon, we recall basic features of SaS
parametrizations of PDF of the virtual photon. We refer the reader
to the web version of this contribution for the figures
illustrating the following observations.
\begin{itemize}
\item Both VDM and pointlike parts drop with increasing $P^2$,
but VDM parts drop much faster.
\item
With increasing $P^2$ the realative importance of VDM parts of both
quark and gluon distribution functions decreases rapidly. For
$M^2\ge 25$ GeV$^2$ the VDM parts of both SaS1D and SaS2D
parametrizations become practically negligible already at
$P^2\approx 3$ GeV$^2$, except in the region of very small
$x\lesssim 0.01$. Hence, also the ambiguity in the separation
(\ref{separation}) is largely irrelevant in this region.
\item
The general pattern of scaling violations remains the same as for
the real photon, except for a subtle difference, best visible (see
Fig. \ref{scaledependence}) when comparing the rescaled PDF for
$P^2=0$ with those at $P^2=1,3$ GeV$^2$. While for $P^2=0$ the
SaSD1 parametrizations of quark distribution functions soften with
increasing $M^2$ and intersect the splitting term at $x\simeq
0.65$, for $P^2\ge 1$ GeV$^2$ they rise with $M^2$ for all $x$ and
stay above the splitting term. These properties reflect the fact
that SaS parametrizations of PDF of the virtual photon do not
satisfy the same evolution equations as PDF of the real one.
\item Pointlike quarks dominate
$D_{\mathrm eff}(x,P^2,M^2)$ at large $x$, while for $x\lesssim
0.5$, most of the pointlike contribution comes from the pointlike
gluons. In particular, the excess of the pointlike contributions to
$D_{\mathrm eff}$ over the contribution of the splitting term,
observed at $x\lesssim 0.5$, comes almost entirely from the
pointlike gluons!
\item For $x\gtrsim 0.6$
the full results are below those given by the splitting term
(\ref{splitterm}) with $M_0^2=P^2$ and one therefore expects the
sum of subtracted direct and resolved contributions to jet
cross--sections to be smaller than the results of unsubtracted
direct calculations.
\end{itemize}
Jet production in ep collisions at HERA in the region of photon
virtualities $P^2\gtrsim 1$ thus offers a promising opportunity for
the identification of nontrivial aspects of PDF of virtual photons
at both small (but not very small) and large values of $x$. All
these conclusions were obtained within the simple LO formalism, but
as shown in the next Section they hold within the framework of NLO
parton level calculations as well.

\subsection{Should we care about ${\mathbf \gamma^*_{L}}$?}
Most of the existing phenomenological analyses of the properties
and interactions of virtual photons as well as all available
parametrizations of their PDF concern transverse photons only.
Neglecting longitudinal photons is a good approximation for
$y\rightarrow 1$, where the flux $f_L^{\gamma}(y,P^2)\rightarrow
0$, as well as for very small virtualities $P^2$, where PDF of
$\gamma_L^*$ vanish by gauge invariance. But how small is ``very
small'' in fact? For instance, should we take into account the
contribution of $\gamma^*_{L}$ to jet cross--section in the region
$E_T^{\mathrm {jet}}\gtrsim 5$ GeV, $P^2\gtrsim 1$ GeV$^2$, where
most of the data on virtual photons extracted from ep collisions at
HERA come from? Despite the fact that the structure of longitudinal
photons has not yet been investigated experimentally and is also
poorly known theoretically, simple parton model estimates (see
\cite{factor} for details) of their effects suggest that in the
mentioned kinematical region $\gamma^*_{L}$ must be taken into
account in the resolved photon contribution but may be safely
neglected in the direct one. This difference comes from the fact
that at small $P^2$ the contributions of $\gamma_L^*L$ to physical
cross--sections behave as $P^2/\hat{s}$ (i.e. vanish for fixed $P^2
$ when $\hat{s}\rightarrow \infty$) in the direct channel, but as
$P^2/\mu^2$ (with $\mu$ a fixed parameter) in the resolved part. In
simple QED based considerations $\mu$ is given by quark masses,
while in realistic QCD we expect it to be given by some
nonperturbative parameter of the order of $1$ GeV.
\begin{figure}\unitlength 1mm
\begin{picture}(160,100)
\put(0,0){\epsfig{file=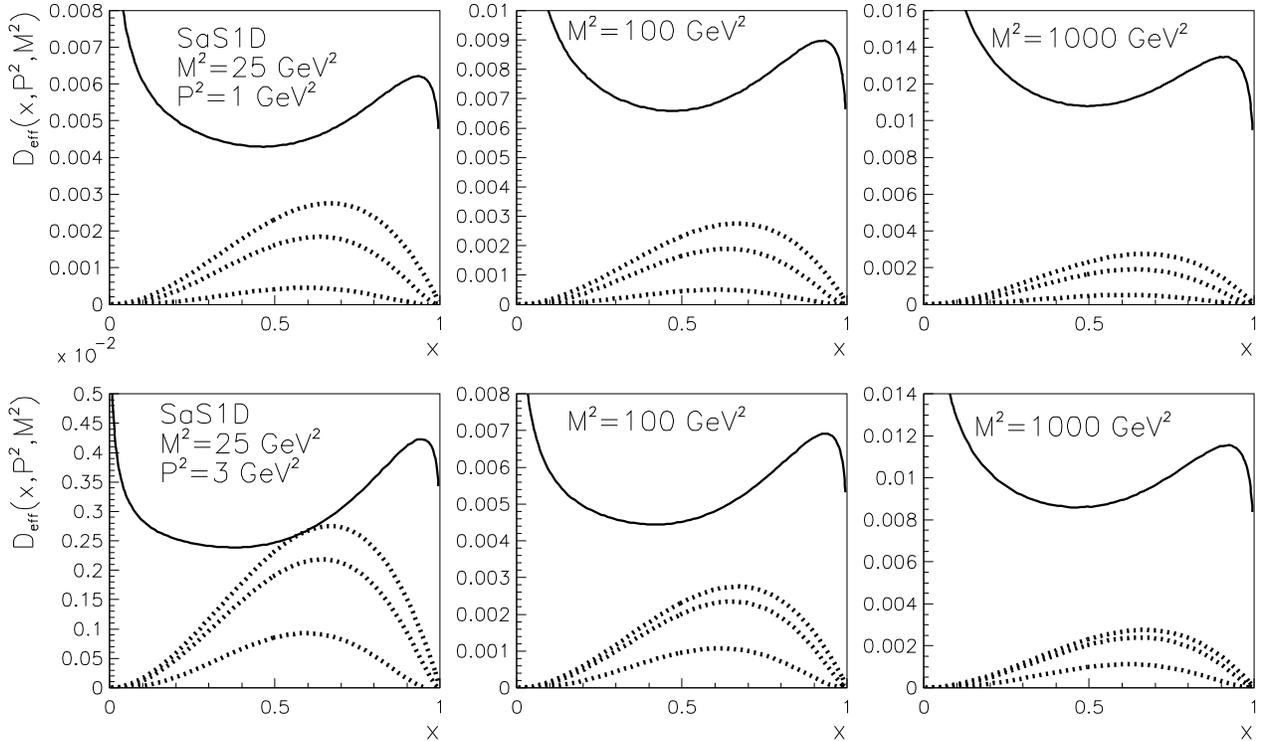,width=\textwidth}}
\end{picture}
\caption{\em
$D_{\mathrm eff}(x,P^2,M^2)$
calculated from SaS1D parametrizations for $\gamma_T^*$ (solid
curves) compared to results for $\gamma_L^*$ displayed by dotted curves
and corresponding from above to $m^2=0,0.1$ and $1$ GeV$^2$.}
\label{f2defftl}
\end{figure}
To assess the importance of $\gamma_L^*$, LO expressions for
$D_{\mathrm eff}(x,P^2,M^2)$ evaluated with SaS1D parametrizations
for $\gamma_T^*$, are compared in Fig. \ref{f2defftl} with the
formula (44) of \cite{factor} for the quark distribution function
of $\gamma_L^*$, treating $m$ in the latter as a free parameter.
The dotted curves in this figure correspond (from below) to $m^2=1$
GeV$^2$, $m^2=0.1$ GeV$^2$ and to the asymptotic expression
$(\alpha/2\pi)12e_q^2x(1-x)$, obtained in the limit $m\rightarrow
0$. As expected, the importance of $\gamma^*_{L}$ depends
sensitively on $m$. Moreover, its contributions relative to those
of $\gamma_T^*$ peak at about $x\approx 0.65$, drop with increasing
$M^2$ (for $P^2$ and $m^2$ fixed), and with increasing $m^2$ (for
$P^2$ and $M^2$ fixed). Fig. \ref{f2defftl} suggests that in the
kinematical region accessible at HERA the contributions of
$\gamma_L^*$ should be taken into account in phenomenological
analyses of data on virtual photon interactions. Better theoretical
knowledge of the structure of $\gamma_L^*$ is clearly needed.

In all considerations of this subsection we simply added the
contributions of $\gamma_T^*$ and $\gamma_L^*$, despite the fact
that the respective fluxes
$f_T^{\gamma}(y,P^2),~f_L^{\gamma}(y,P^2)$ differ. Our conclusions
are therefore directly relevant for small values of $y$ only
\footnote{Note, however, that even at
$y=0.4$ the ratio $f_T^{\gamma}(y,P^2)/f_L^{\gamma}(y,P^2)=1.16$ is
still quite close to $1$.}, but it is trivial to modify the above
considerations by taking the respective fluxes properly into
account.

\section{PDF of the virtual photon in NLO QCD calculations}
In \cite{H1eff} data on dijet production in the region of photon
virtualities $1\le P^2\le50$ GeV$^2$, and for jet transverse
energies $E_T^{\mathrm jet}\ge 5$ GeV have been analyzed within the
framework of effective PDF defined in (\ref{deff}). This analysis
shows that in the kinematical range $1~{\mathrm {GeV}}^2\lesssim
P^2\lesssim E_T^2$ the data agree reasonably with the expectations
based on SaS parametrizations of PDF of the virtual photon. The
same data may, however, be also analyzed using the NLO parton level
Monte--Carlo programs that do not introduce the concept of PDF of
virtual photons, like DISENT, MEPJET or DISASTER$++$. Nevertheless,
so long as $P^2\ll M^2\approx E_T^2$, the pointlike parts of PDF
incorporate numerically important effects of a part of higher order
corrections, namely those coming from collinear emission of partons
in Fig. \ref{figpl}. To illustrate this point we shall now discuss
dijet cross--sections calculated by means of JETVIP \cite{JETVIP},
currently the only NLO parton level Monte--Carlo program that
includes both the direct and resolved photon contributions.

All the above mentioned parton level NLO MC programs contain the
same full set of partonic cross--sections
\footnote{In this subsection the various terms considered are
characterized by the powers of $\alpha$ and $\alpha_s$ appearing in
hard scattering cross--sections. Writing $O(\alpha^j\alpha_s^k)$
will thus mean terms proportional to $\alpha^j\alpha_s^k$, {\em
not} terms {\em up to} this order! For approximations taking into
account the first two or three powers of $\alpha_s$, in either
direct or resolved channel, the denomination NLO and NNLO are
used.} for the direct photon contribution up the order
$\alpha\alpha_s^2$. In addition JETVIP includes also the resolved
photon contribution. Once the concept of virtual photon structure
is introduced, part of the direct photon contribution, namely the
splitting term (\ref{splitterm}), is subtracted from it and
included in PDF appearing in the resolved photon contribution. To
avoid confusion we shall use the term ``direct unsubtracted''
(DIR$_{\mathrm {uns}}$) to denote complete NLO direct photon
contribution and reserve the term ``direct'' (DIR) for the results
after this subtraction. In this terminology the complete NLO
calculations is given by the sum of direct and resolved parts and
denoted DIR$+$RES.

For complete $O(\alpha_s^2)$ calculations only the LO resolved
photon contribution must be added to the $O(\alpha\alpha_s^2)$
direct one. However, for reasons outlined in the next subsection,
JETVIP includes also NLO resolved contributions. This might seem
inconsistent as the corresponding $O(\alpha\alpha_s^3)$ direct
photon terms are not yet available and thus not included.
Nevertheless, this procedure makes sense precisely because of a
clear physical meaning of PDF of the virtual photon!

\subsection{Factorization mechanism in $\gamma$p interactions}
The main argument for adding $O(\alpha_s^3)$ resolved photon terms
to $O(\alpha\alpha_s^2)$ direct and $O(\alpha_s^2)$ resolved photon
contributions is based on specific way factorization mechanism
works for processes involving initial photons. This point is
crucial but subtler and we therefore merely summarize the
conclusions and refer the reader to the website version of this
contribution, or to \cite{factor}, for details. In the absence of
$O(\alpha\alpha_s^3)$ direct calculations, we have two options:
\begin{itemize}
\item To stay within the framework of complete $O(\alpha_s^2)$
calculations, including the LO resolved and NLO direct
contributions, but with no mechanism for the cancellation of
the dependence of PDF of the virtual photon on the factorization
scale $M$.
\item To add to the previous framework the $O(\alpha_s^3)$ resolved
photon contributions, which provide necessary cancellation
mechanism with respect to the part of factorization scale
dependence of photonic PDF ``generated'' by the homogeneous part of
the evolution equations (\ref{Sigmaevolution}--\ref{NSevolution}).
The drawback of this procedure is the fact that the NLO resolved
photon terms do not represent a complete set of $O(\alpha_s^3)$
contributions.
\end{itemize}
In our view the second strategy, adopted in JETVIP, is more
appropriate. In fact one can look at $O(\alpha_s^3)$ resolved
photon terms as results of approximate evaluation of the so far
uncalculated $O(\alpha\alpha_s^3)$ direct photon diagrams in the
collinear kinematics. There are of course $O(\alpha\alpha_s^3)$
direct photon contributions
that cannot be obtained in this way, but
we are convinced that it makes sense to build phenomenology on this
framework.

Note that for the $O(\alpha_s^2)$ resolved terms the so far unknown
$O(\alpha\alpha_s^3)$ direct photon contributions provide the first
chance to generate pointlike gluons inside the photon. To get the
gluon in $O(\alpha_s^3)$ resolved photon contributions would
require evaluating diagrams of even higher order
$O(\alpha\alpha_s^4)$! In summary, although the pointlike parts of
quark and gluon distribution functions of the virtual photon are in
principle included in higher order perturbative corrections and can
therefore be considered as expressions of ``interactions'' rather
than ``structure'', their uniqueness and phenomenological
usefulness definitely warrant their introduction as well as their
names.

\subsection{Dijet production at HERA}
To make our conclusions potentially relevant for analyses of HERA
data on jet produnction we have chosen the following kinematical
region (jets with highest and second highest $E_T$ are labelled
``1'' and ``2'')
\begin{description}
\item{\bf Jet transverse energies:} asymmetric cuts~~~$E_T^{(1)}\ge 7$ GeV,
$E_T^{(2)}\ge 5$ GeV;
\item{\bf Photon virtuality:} four windows in $P^2$
$$1.4\le P^2\le 2.4~{\mathrm {GeV}}^2;
~~2.4\le P^2\le 4.4~{\mathrm {GeV}}^2;~~
4.4\le P^2\le 10~{\mathrm GeV}^2;~~10\le P^2\le 25~{\mathrm {GeV}}^2
    $$
\item{\bf Jet pseudorapidities} in $\gamma^*$ CMS:
 $-2.5 \le \eta^{(i)}\le 0,~i=1,2$
\end{description}
The cuts were chosen in such a way that throughout the region
$P^2\ll E^2_T$, thereby ensuring that the virtual photon lives long
enough for its ``structure'' to develop before the hard scattering
takes place. As far as jet transverse energies are concerned, we
have chosen the asymmetric cut scenario: $E_T^{(1)}\ge
E_T^{c}+\Delta,~E_T^{(2)}\ge E_T^{c}$, which avoids the problems
\cite{FR} coming from the region where $E_T^{(1)}\approx
E_T^{(2)}$. The asymmetric cut option is appropriate if one plots
separately the $E_T^{(1)}$ and $E_T^{(2)}$ distributions, or, as we
do, their sum, called distribution of ``trigger jets''
\cite{JETVIP}. To determine the value of $\Delta$ optimally, we
evaluated the integral $\sigma(\Delta)$ over the selected region in
$E_T^{(1)}-E_T^{(2)}$ plane as a function of $\Delta$ and on the
basis thereof took $\Delta=2$ GeV for all $P^2$.

In our analysis jets are defined by means of the cone
algorithm. At NLO parton level all jet algorithms are essentially
equivalent to the cone one, supplemented with the parameter
$R_{\mathrm sep}$, introduced in order to bridge the gap between
the application of the cone algorithm to NLO parton level
calculations and to hadronic systems (from data or MC), where one
encounters ambiguities concerning the seed selection and jet
merging. In a general cone algorithm two objects (partons, hadrons
or calorimetric cells) belong to a jet if they are within the
distance $R$ from the jet center. Their relative distance
satisfies, however, a weaker condition
\begin{equation}
\Delta R_{ij}=\sqrt{(\Delta \eta_{ij})^2+(\Delta \phi_{ij})^2}
\le \frac{E_{T_i}+E_{T_j}}{{\mathrm max}(E_{T_i},E_{T_j})}R.
\end{equation}
The parameter $R_{\mathrm sep}$ governs the maximal distance
between two partons within a single jet, i.e. two partons form a
jet only if their relative distance $\Delta R_{ij}$ satisfies
the condition
\begin{equation}
\Delta R_{ij}\le {\mathrm min}\left[
\frac{E_{T_i}+E_{T_j}}{{\mathrm max}(E_{T_i},E_{T_j})}R,
R_{\mathrm sep}\right].
\end{equation}
The question which value of $R_{\mathrm sep}$ to choose for the
comparison of NLO parton level calculations with the results of the
cone algorithm applied at the hadron level is nontrivial and we shall
therefore present JETVIP
results for both extreme choices $R_{\mathrm
sep}=R$ and $R_{\mathrm {sep}}=2R$. To define momenta of jets
JETVIP uses the standard $E_T$--weighting recombination procedure,
which leads to massless jets. To assess the reliability of our
conclusions we have furthermore
investigated the following uncertainties:
\begin{description}
\item {\bf Choice of PDF:} We have taken CTEQ4M and SAS1D sets of
PDF of the proton and photon respectively as our canonical choice.
Both of these sets treat quarks, including $c$ and $b$ ones, as
massless above their respective mass thresholds, as required by
JETVIP, which uses LO and NLO matrix elements of massless partons.
Taking into account our cuts on jet transverse energies, we set
$N_f=4$ in any calculations discussed below. PDF of the proton are
fairly well determined from global analyses of CTEQ and MRS groups
and we have therefore estimated the residual uncertaintly related
to the choice of PDF of the proton by comparing the CTEQ4M results
to those obtained with MRS(2R) set. The differences are very small,
between 1.5\% at $\eta=0$ and 3\% at $\eta=-2.5$.
\item {\bf Factorization scale dependence:} In principle proton and
(in resolved channel) photon are associated with different
factorization scales $M_p$ and $M_{\gamma}$, but we followed the
standard practice and set $M_p=M_{\gamma}=M=\kappa
(E_T^{(1)}+E_T^{(2)})/2$. The factorization scale dependence was
estimated by performing the calculations for $\kappa=1/2,1,2$.
\item {\bf Renormalization scale dependence:} The dependence of
perturbative calculations on the renormalization scale $\mu$ is in
principle a separate ambiguity, unrelated to that of factorization
scale $M$, but we followed the common practice and set $\mu=M$.
\item {\bf Jet algorithm ambiguities:} To see how much our conclusions
depend on $R_{\mathrm sep}$ we performed our calculations for two
extremes values: $R_{\mathrm sep}=R,2R$.
\item {\bf Hadronization corrections:} For any comparison of
parton level calculations with experimental data understanding
these corrections is crucial, but the problem is particularly
pressing for jets with moderate $E_T$, like those used in all
analyses of virtual photon structure. Hadronization corrections are
not simple to define, but adopting the definition used by
experimentalists \cite{hadkor}, we have found that they depended
sensitively and in a correlated manner on transverse energies and
pseudorapidities of jets. For $E_T^c=5$ GeV they start to rise
steeply below $\eta\doteq -2.5$ and we therefore required both jets
to lie in the region $-2.5\le\eta^{(i)}\le 0$, where hadronization
corrections are flat in $\eta$ and do not exceed $20$\%.
\end{description}
\begin{figure}\unitlength 1mm
\begin{picture}(160,80)
\put(0,0){\epsfig{file=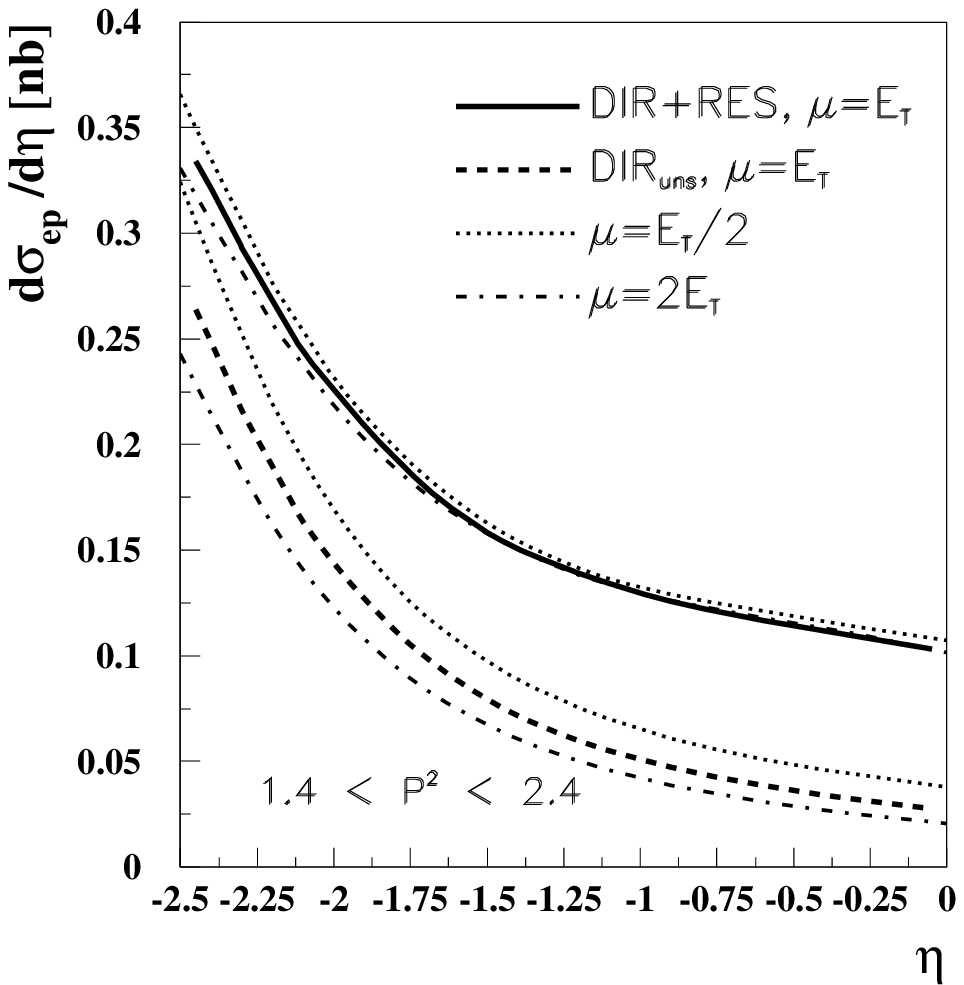,width=7.5cm}}
\put(80,0){\epsfig{file=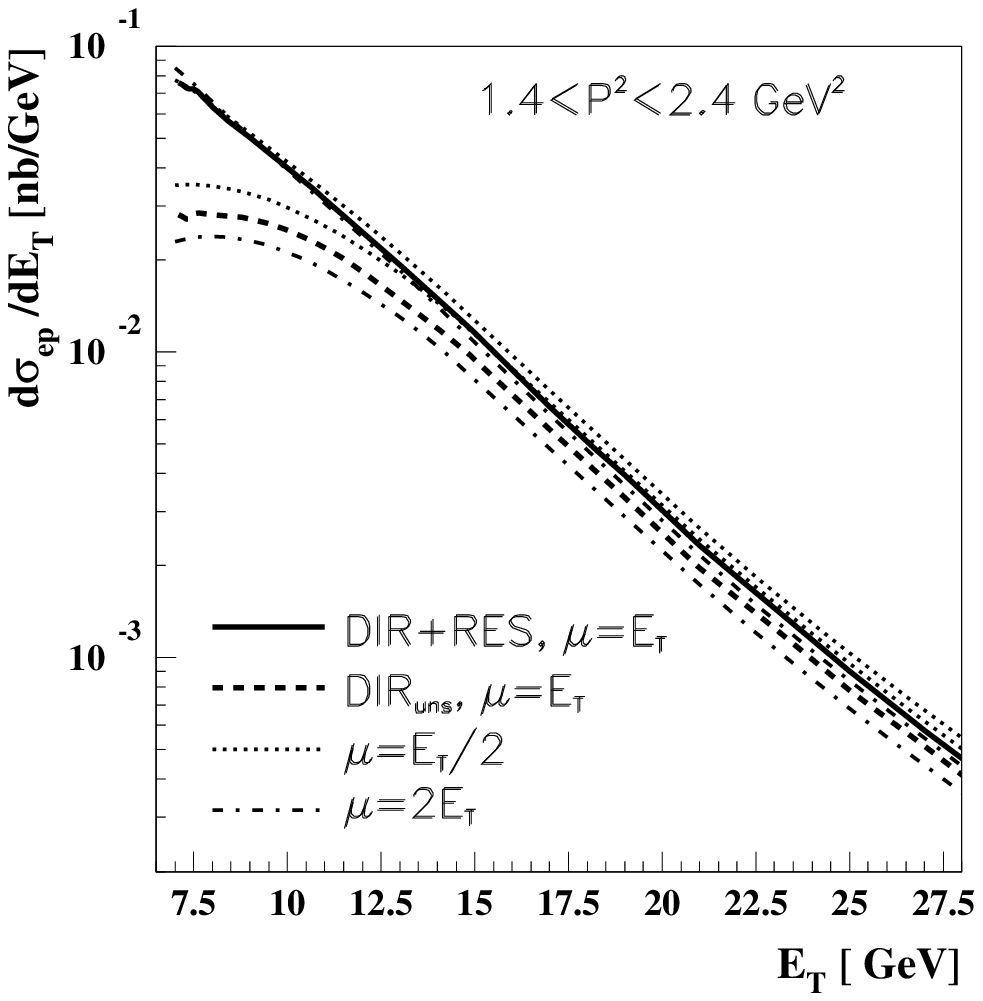,width=7.5 cm}}
\end{picture}
\caption{\em Scale dependence of the distributions
${\mathrm d}\sigma/{\mathrm d}\eta$ and ${\mathrm d}\sigma/{\mathrm
d}E_T$ at the NLO. All curves correspond to $R_{\mathrm {sep}}=2R$.
The dotted and dashed--dotted curves have the same meaning for
DIR$+$RES as well as DIR$_{\mathrm {uns}}$ calculations.}
\label{win1}
\end{figure}

\begin{figure}
\epsfig{file=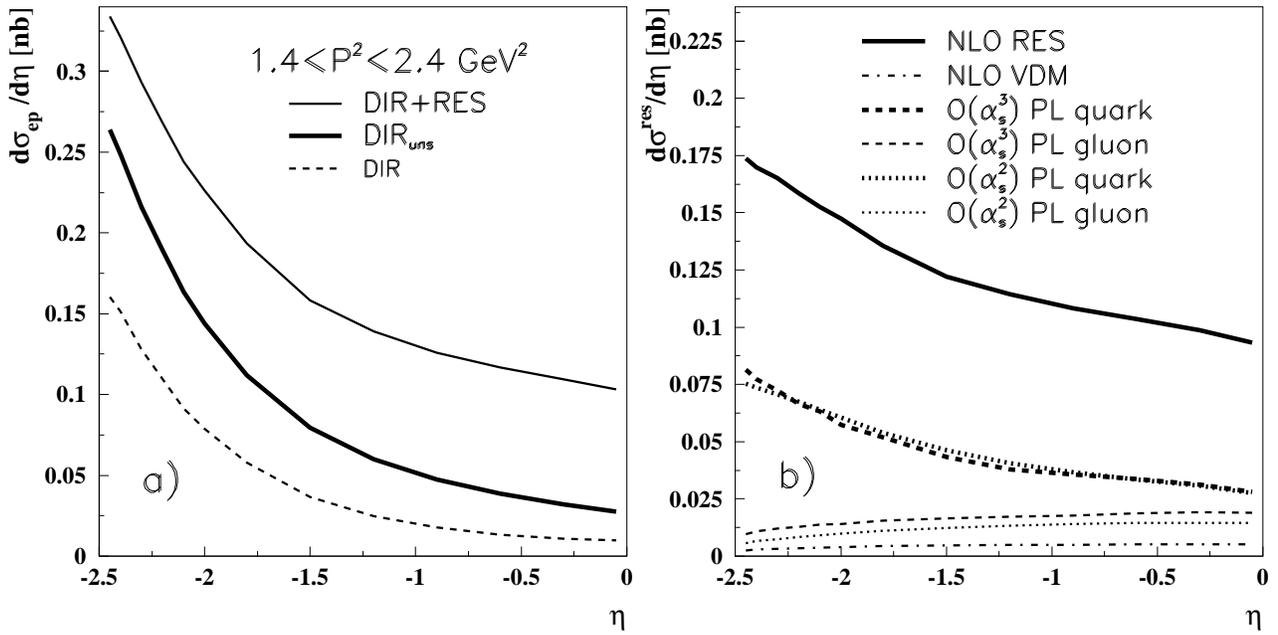,width=\textwidth}
\caption{\em Comparison of DIR$+$RES, DIR$_{\mathrm uns}$ and
DIR results for ${\mathrm d}\sigma/{\mathrm d}\eta$ (a). In b)
individual contributions to ${\mathrm d}\sigma^{\mathrm
res}_{\mathrm {NLO}}/{\mathrm d}\eta$ are plotted.}
\label{reseta1}
\end{figure}

\begin{figure}\unitlength 1mm
\begin{picture}(160,60)
\put(0,0){\epsfig{file=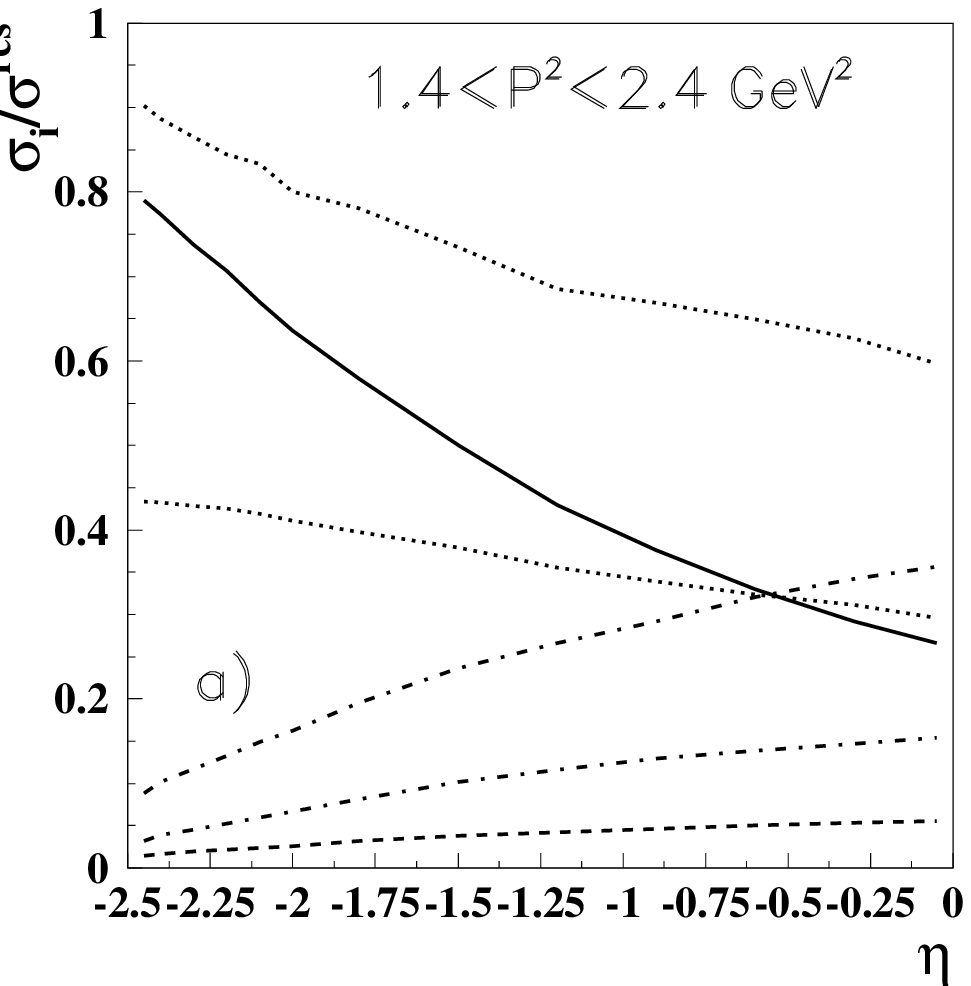,width=5.5cm}}
\put(55,0){\epsfig{file=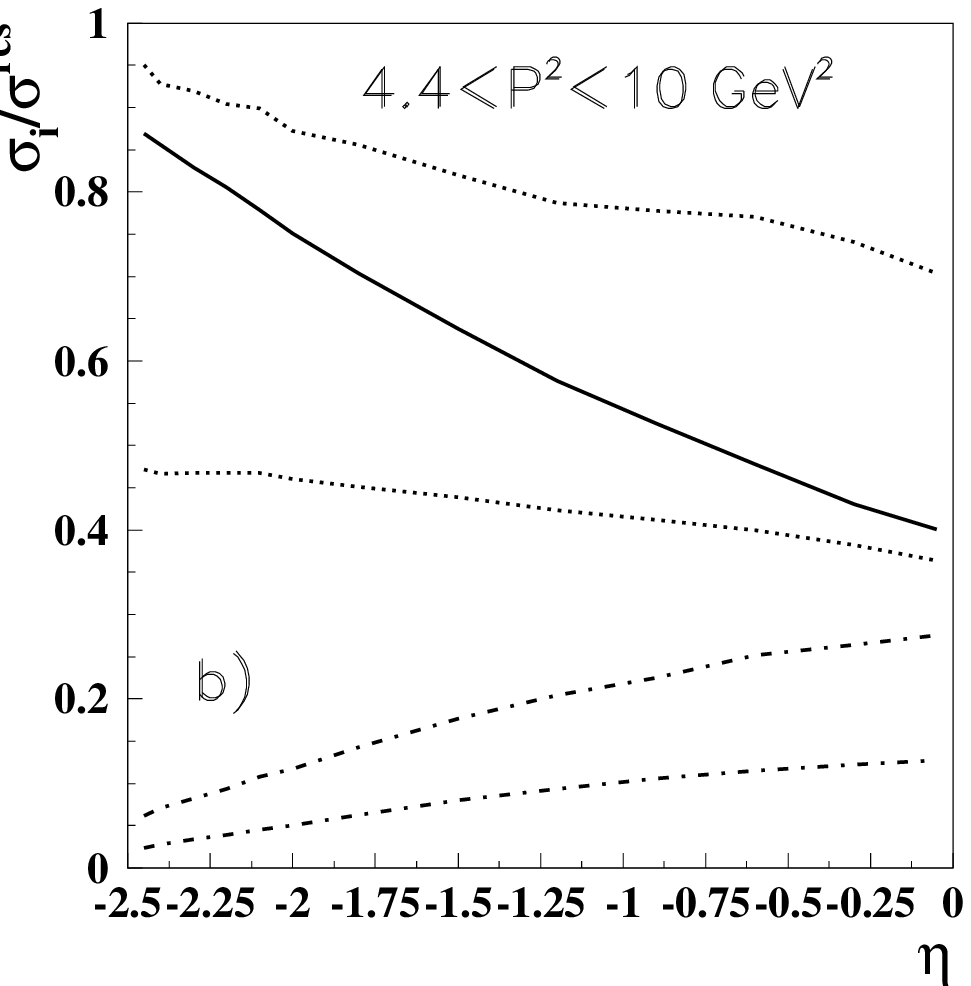,width=5.5cm}}
\put(110,0){\epsfig{file=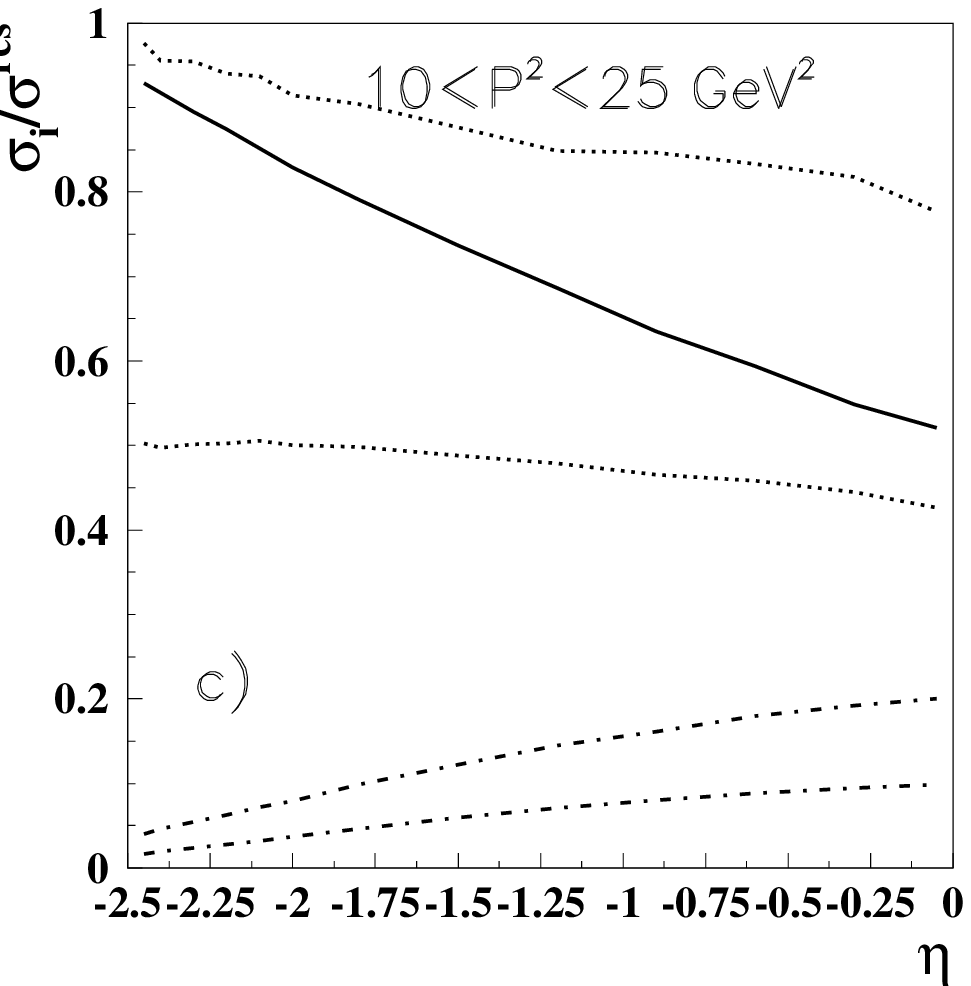,width=5.5cm}}
\end{picture}
\caption{\em Fractional contributions to
$\sigma^{\mathrm {res}}$. Upper and lower dotted (dashed--dotted)
curves correspond to pointlike quarks (gluons) convoluted with NLO
and LO partonic cross--sections. The lower dashed curve in a) gives
the NLO VDM contribution. The solid curves show the ratia of NLO
DIR$_{\mathrm {uns}}$ calculations to the NLO DIR$+$RES ones.}
\label{fractions}
\end{figure}

\begin{figure}\centering
\epsfig{file=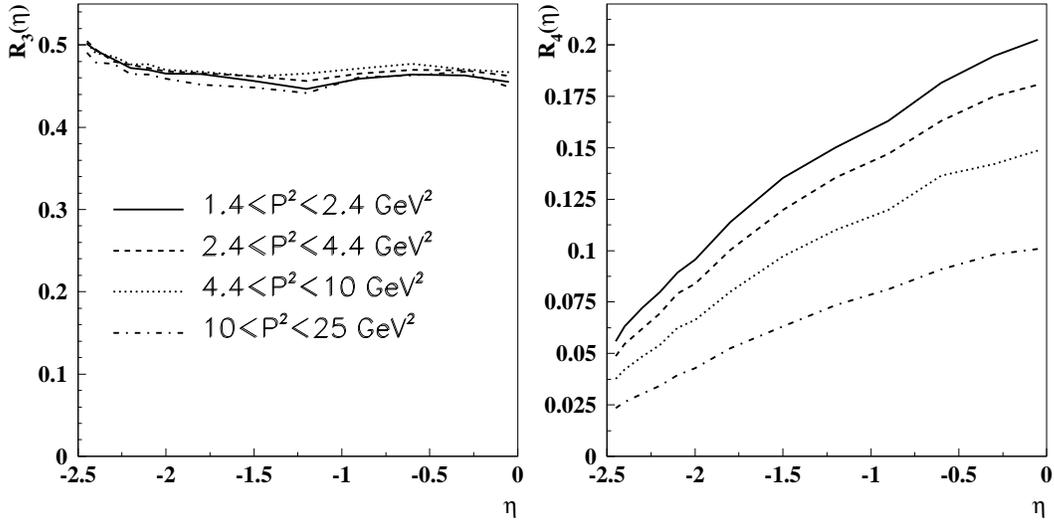,width=14cm}
\caption{\em Nontriviality fractions $R_3$ and $R_4$ as functions of
$\eta$ and $P^2$.}
\label{nontrivial}
\end{figure}

For virtual photons JETVIP can be run in two different modes:
\begin{description}
\item{\bf DIR$_{\mathrm {uns}}$:} only the NLO unsubtracted direct
photon calculations are performed without introducing the concept
of virtual photon structure.
\item{\bf DIR$+$RES:}  employs the
concept of PDF of the virtual photon and gives jet cross--sections
as sums DIR$+$RES of subtracted direct and resolved photon
contributions.
\end{description}
We first discuss results for the first window $1.4\le P^2\le 2.4$
GeV$^2$. In Fig. \ref{win1} ${\mathrm {d}}\sigma/{\mathrm {d}}\eta$
and ${\mathrm {d}}\sigma/{\mathrm {d}}E_T$ distributions of trigger
jets obtained within the DIR$+$RES approach are compared to those
of DIR$_{\mathrm {uns}}$ one. All curves correspond to $R_{\mathrm
sep}=2$. The difference between the results of these two approaches
is significant in the whole range of $\eta$, but particularly large
close to the upper edge $\eta=0$, where the DIR$+$RES results
exceed the DIR$_{\mathrm {uns}}$ ones by a factor of more than 3!
In ${\mathrm {d}}\sigma/{\mathrm {d}}E_T$ distributions this
difference comes predominantly from the region of $E_T$ close to
the lower cut--off $E_T^c+\Delta=7$ GeV. Fig. \ref{win1} also shows
that the scale dependence is nonnegligible in both approaches, but
does not invalidate the main conclusion drawn from this comparison.
The dependence of the above results on $R_{\mathrm sep}$ (not
shown) is almost imperceptible for DIR$_{\mathrm {uns}}$
calculations and below $10$\% for the DIR$+$RES ones. To track down
the origins of the observed large differences between DIR$+$RES and
DIR$_{\mathrm {uns}}$ results, we did two simple exercises. In Fig.
\ref{reseta1}a the DIR$+$RES and DIR$_{\mathrm {uns}}$ results are
compared to the subtracted direct (DIR) ones. The difference
between the DIR$+$RES and DIR curves, defining the resolved photon
contribution ${\mathrm d}\sigma^{\mathrm res}/{\mathrm d}\eta$, is
then split into the contributions of:
\begin{itemize}
\item the VDM part of photonic PDF convoluted with complete NLO
(i.e. up to the order $O(\alpha_s^3)$) parton level cross--sections
(denoted NLO VDM),
\item the pointlike quark and gluon distribution functions convoluted
with $O(\alpha_s^2)$ and $O(\alpha_s^3)$ parton level
cross--sections
\end{itemize}
and plotted in Fig. \ref{reseta1}b. Fractional contributions of LO
(i.e. $O(\alpha_s^2)$) and NLO (i.e. $O(\alpha_s^2)+O(\alpha_s^3)$)
terms to $\sigma^{\mathrm {res}}$ are plotted in Fig.
\ref{fractions}a as functions of $\eta$. Several conclusions can be
drawn from Figs. \ref{reseta1}--\ref{fractions}:
\begin{itemize}
\item The contribution of the VDM part of photonic PDF is very small
and perceptible only close to $\eta=0$.
Integrally it amounts to about 3\%. Using SaS2D
parametrizations would roughly double this number.
\item The inclusion of $O(\alpha_s^3)$ resolved photon contributions
is numerically important in the whole range $-2.5\le
\eta\le 0$. Interestingly, throughout this interval the
$O(\alpha_s^3)$ results, particularly those of pointlike quarks,
are close to the $O(\alpha_s^2)$ ones.
\item At both $O(\alpha_s^2)$ and $O(\alpha_s^3)$ orders pointlike
quarks dominate ${\mathrm d}\sigma^{\mathrm res}/{\mathrm d}\eta$
at large negative $\eta$, whereas
as $\eta\rightarrow 0$ the fraction of ${\mathrm d}\sigma^{\mathrm
res}/{\mathrm d}\eta$ coming from pointlike gluons increases
towards about $40$\% at $\eta=0$.
\end{itemize}
We reiterate that pointlike gluons carry nontrivial information
already in convolutions with $O(\alpha_s^2)$ partonic
cross--sections because in unsubtracted direct calculations such
contributions appear first at the so far uncalculated order
$\alpha\alpha_s^3$. The results of pointlike quarks convoluted with
$O(\alpha_s^3)$ partonic cross--sections would be included in
unsubtracted direct calculations starting at the order
$\alpha\alpha_s^3$, whereas for pointlike gluons this would require
evaluation of unsubtracted direct terms of even higher order
$\alpha\alpha_s^4$! In JETVIP the nontrivial aspects of taking into
account the resolved photon contributions can be characterized
\footnote{Disregarding the VDM part of resolved contribution
which is tiny in our region of photon virtualities.} by the
``nontriviality fractions'' $R_3$, $R_4$
\begin{equation}
R_3\equiv
\frac{q^{\mathrm {PL}}\otimes\sigma_q^{\mathrm {res}}(O(\alpha_s^3))+
      G^{\mathrm {PL}}\otimes\sigma_G^{\mathrm {res}}(O(\alpha_s^2))}
      {\sigma^{\mathrm {res}}},~~~
R_4\equiv
\frac{G^{\mathrm {PL}}\otimes\sigma_G^{\mathrm {res}}(O(\alpha_s^3))}
    {\sigma^{\mathrm {res}}},
\label{nontr}
\end{equation}
plotted as functions of $\eta$ and $P^2$ in Fig. \ref{nontrivial}.
Note that at $\eta=0$ almost 70\% of $\sigma^{\mathrm {res}}$ comes
from these origins. This fraction rises even further in the region
$\eta>0$.

So far we have discussed the situation in the first window $1.4\le
P^2\le 2.4$ GeV$^2$. As $P^2$ increases the patterns of scale and
$R_{\mathrm {sep}}$ dependences change very little. On the other
hand, the fractions plotted in Fig. \ref{fractions} and
\ref{nontrivial} vary noticably:
\begin{itemize}
\item The unsubtracted direct photon contributions
(DIR$_{\mathrm {uns}}$) represent increasing fractions of the full
NLO JETVIP results.
\item The relative contribution of pointlike gluons with respect to
pointlike quarks decreases.
\item The nontriviality factor $R_4$ (which comes entirely from
pointlike gluons) decreases, whereas $R_3$, which is dominated by
pointlike quarks and flat in $\eta$, is almost independent of
$P^2$.
\end{itemize}
All these features of JETVIP calculations reflect the fundamental
fact that as $P^2$ rises towards the factorizations scale
$M^2\approx E_T^2$ the higher order effects incorporated in
pointlike parts of photonic PDF vanish and consequently the
unsubtracted direct results approach the DIR$+$RES ones. The
crucial point is that for pointlike quarks and gluons this approach
is governed by the ratio of $P^2/M^2$. The nontrivial effects
included in PDF of the virtual photon thus persist for arbitrarily
large $P^2$, provided we stay in the region where $P^2\ll M^2$.

\section{Summary and conclusions}
We have analyzed the content of parton distribution functions of
the virtual photon within the framework of SaS approach to their
parametrization. In this approach, quark as well as the gluon
distribution functions can be separated into the nonperturbative
VDM and pointlike parts, the latter being in principle calculable
by perturbative means. The inherent ambiguity of this separation,
numerically large for the real photon, becomes phenomenologically
irrelevant for virtual photons with $P^2\gtrsim 3$ GeV$^2$. In this
region quark and gluon distribution functions of the virtual photon
are dominated by their (reasonably unique) pointlike parts, which
have clear physical origins. We have analyzed the nontrivial
aspects of these pointlike distribution functions and, in
particular, pointed out the role of pointlike gluons in leading
order calculations of jet cross--section at HERA.

At the NLO we have found a significant difference between JETVIP
results in approaches with and without the concept of virtual
photon structure. While for the real photon analogous difference is
in part ascribed to the VDM components of photonic PDF, for
moderately virtual photons it comes mostly from pointlike quarks
and pointlike gluons. Although their contributions are in principle
contained in higher order direct photon calculations, to get them
in practice would require calculating at least
$O(\alpha\alpha_s^3)$ and $O(\alpha\alpha_s^4)$ unsubtracted direct
contributions. In the absence of such calculations the concept of
PDF of the virtual photon is therefore phenomenologically
indispensable.

We have also shown that despite the expected dominance of the
transversely polarized photons, the longitudinal polarization of
virtual photons may play a numerically important role in the region
of moderate virtualities accessible at HERA and should therefore be
taken into account in analyses of relevant experimental data.

\noindent
{\large \bf Acknowledgments:} We are grateful to G. Kramer and B.
P\"{o}tter for discussions on the interactions of virtual photons
and to B. P\"{o}tter for help with running JETVIP.

\end{document}